\documentclass{aa}  

\usepackage{graphicx}
\usepackage{txfonts}
\usepackage{lipsum}
\usepackage{subcaption}         
\usepackage{lscape}             
                                 
\usepackage{placeins}          
                            
\usepackage{multicol}

\usepackage{color}

\begin{document}

   \title{Deep learning from the crowd}

   \subtitle{Fundamentals of morphological galaxy classification}

   \author{Luis Enrique Sucar\textsuperscript{1,2}\fnmsep\thanks{Corresponding author.} \and Carlos del Burgo\textsuperscript{1,3} 
   \and Jonathan Serrano-P\'erez\textsuperscript{2,4}}

   \institute{
   Instituto de Astrof\'\i sica de Canarias, V\'\i a L\'actea S/N, La Laguna 38200, Tenerife, Spain 
   \and Instituto Nacional de Astrof\'isica, \'Optica y Electr\'onica, Luis Enrique Erro \#1, CP 72840, Tonantzintla, Puebla, Mexico
   \and Departamento de Astrof\'\i sica, Universidad de la Laguna, La Laguna 38200, Tenerife, Spain
   \and Facultad de Ciencias B\'asicas, Ing. y Tec., Universidad Aut\'onoma de Tlaxcala, Carretera Apizaquito S/N, CP 90401, San Luis Apizaquito, Tlaxcala, Mexico \\
             \email{esucar@inaoep.mx, cburgo@ull.edu.es, js.perez@inaoep.mx}\\ }
            
   \date{Received April 15, 2026}

  \abstract
  {}
   {The objective of this work is to adapt a deep neural network model to perform galaxy morphological classification 
   trained from crowd annotations, considering the training scheme, the agreement between the annotators, and the hierarchy.} 
   {We use Galaxy Zoo 1 as our experimental testbed and trained a convolutional neural network (CNN) for the automatic classification of galaxies' morphologies. We analyze the impact of the following aspects on the classification accuracy and training efficiency: (i) Training only the last layer vs. training all the network; (ii) Classification with only the CNN vs. considering the hierarchy; (iii) Comparing the models trained with different amounts of data and levels of agreement between the annotators; (iv) 
Training by stages, transferring knowledge from one model to another; and (v) Combining several models as an ensemble. }
   {From the experiments, we derive the following results: (i) Training all the layers in the network significantly improves the accuracy (10\% increase in exact match), compared to training only the last layer; (ii) There is a tradeoff between the amount of data and the level of agreement between the annotators used for training; 
   
   (iii) Using the hierarchy can improve accuracy when the amount of training data is reduced; (iv) Training by stages through transfer learning (curriculum learning) produces higher accuracy for limited data; (v) Ensembles can improve accuracy; (vi) Models achieve a low accuracy for the most difficult cases, but, if we consider hierarchical measures, we can derive useful results for upper levels in the hierarchy. An accuracy above 99\% is achieved when training all layers of the network and considering a  high agreement between the annotators}
   {Training deep learning models from crowd annotations involves additional challenges than learning from  hard annotations. We analyze the impact of different aspects when training a CNN from the crowd for morphological galaxy classification. The conclusions derived from this work can be useful for more complex galaxy classifications.}


   \maketitle
   \nolinenumbers

\section{Introduction}

\begin{figure}[bt]
   \centering
   \includegraphics[width=\hsize]{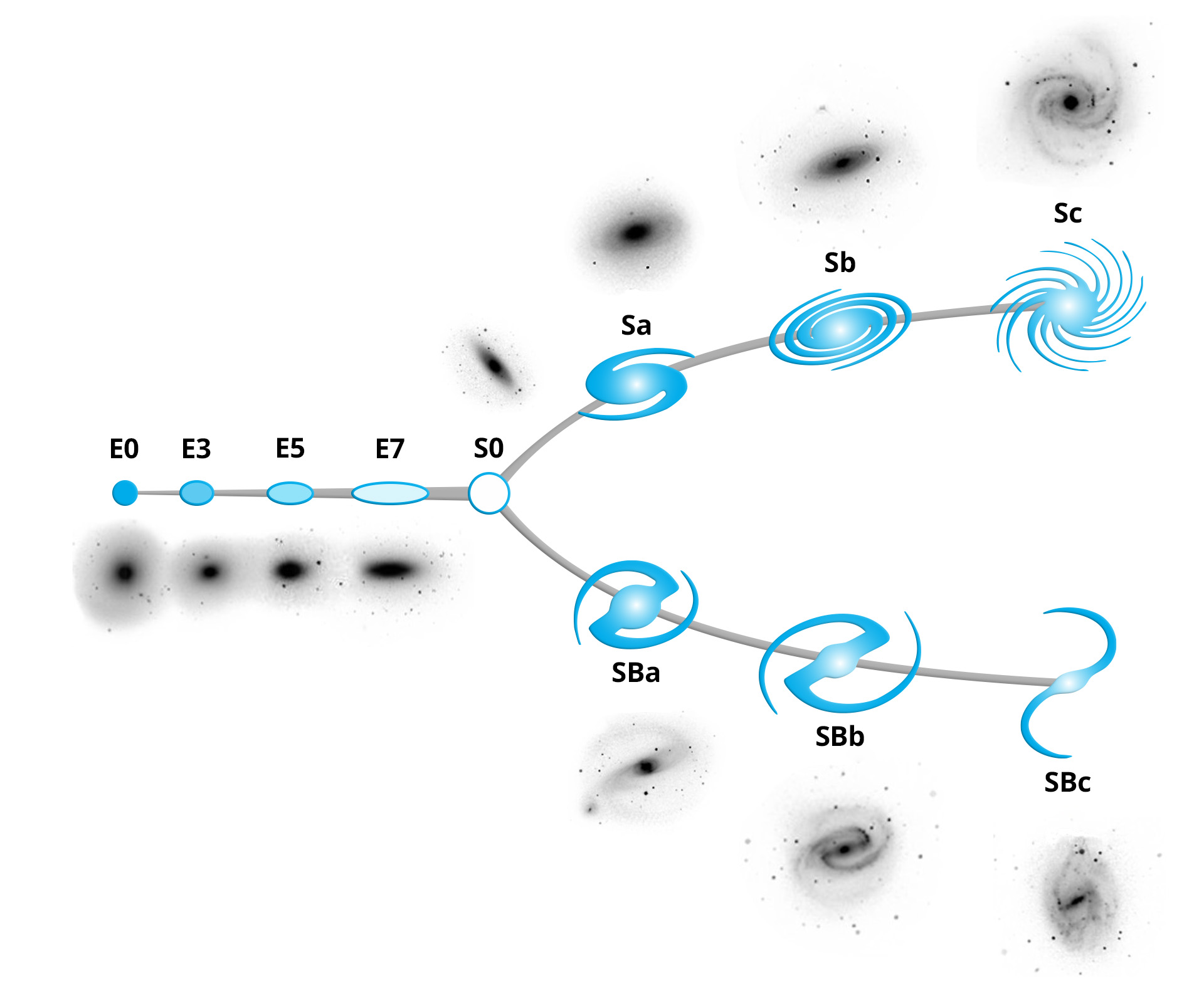}
      \caption{Hubble's  tuning fork scheme for morphological classification of galaxies.}
         \label{hubble}
   \end{figure}

Classifying galaxies by their morphology is an important first step for their analysis, providing important information about their physical properties and  evolution. \cite{Hubble:26} proposed the first classification of galaxies according to the most prominent features of their morphology, which is structured as a  hierarchy divided into three main classes (Hubble's tuning fork): elliptical, spiral and lenticular, all including several subclasses as illustrated in Figure \ref{hubble}. Later, other more refined classification schemes have been proposed, such as the De Vaucouleurs system \citep{Vaucouleurs:59}. Since the seminal work of Hubble, most of the classifications have been done through visual inspection by professional astronomers \citep[e.g.,][]{Sandage:61, Vancoulers:91, Fukugita:07}. However, the advent of large-scale sky surveys makes it almost impossible to classify galaxies by professionals. For instance, the Sloan Digital Sky Survey \citep[SDSS;][]{York:00} has nearly 900,000 galaxies, while the DESI Legacy Image Survey (DESI-LS), which is a combination of three individual surveys \citep{Walmsley:23}, comprises 8.7 million galaxies.

Given the challenge of annotating such large numbers of galaxies, the astrophysics community has relied on training non-experts to label the images through what is known as  citizen science. Through a website, volunteers label each image.  They are given a brief tutorial and asked to identify some `standard' galaxies; those that can correctly classify the initial set proceed to label the galaxy images, which are selected at random from the entire sample. In this way, several 
crowd annotated datasets have been developed, including Galaxy Zoo 1 \citep[GZ1;][]{Lintott:08} and Galaxy Zoo 2 \citep[GZ2;][]{Willett:13} drawn from the SDSS galaxy survey;  Galaxy Zoo DECaLS \citep{Walmsley:22} from the DESI-LS sky survey, Galaxy Zoo CANDELS \citep{Simmons:16} from the CANDELS survey; and Galaxy Zoo Hubble \citep{Willett:16}, with images selected from various publicly released Hubble Space Telescope Legacy programs.

Although these crowd-annotated data sets provide useful results, in particular when there is a high  agreement between annotators, the increasing scale of recent sky surveys (and those to come) makes it impractical to manually annotate all images even by a large number of volunteers. For example, classifying approximately 400,000 galaxy images by 7.5 million volunteers for GZ DECaLS took 4.5 years; at the same rate, collecting 40 classifications per image for the entire DESI-LS would take approximately 200 years \citep{Walmsley:23}. Thus, the task requires automatic classification methods based on machine learning (ML) to annotate such large quantities of galaxy images.

Modern deep learning models are based on multi-layered neural networks, and in particular convolutional neural networks (CNNs) have become the state of the art for classifying galaxies from their morphology alone \citep{Huertas:23}. There are several interrelated issues that could have an important impact when training a deep neural network for galaxy morphological classification. Although there have been some recent works that have trained deep learning networks based on crowd sourcing \citep{Gharat:22, Khramtsov:22, Walmsley:24} for galaxy classification, a comprehensive study of factors influencing model performance has not yet been conducted. This is the motivation of this work, for which we identify three fundamental components:
\begin{description}
\item [Training scheme:] How to train a deep learning model impacts its ability to perform certain tasks and, therefore, its performance, especially with limited or imbalanced 
data. In particular, using a model previously trained on large amounts of data (known as  transfer learning) and which parts of the network to retrain, are critical elements.
\item [Annotators agreement:] Learning from crowd-annotated data is challenging, as there could be uncertainty about the annotations due to several reasons; this is known as learning from disagreement (see Section \ref{disagr}). Thus, this aspect should be taken into account when training a machine learning model. 
\item [Hierarchy:] The hierarchy provides additional information that should be taken into account when developing a system for galaxy classification because of two reasons: (i) it could improve classification performance by considering the restrictions imposed by the hierarchy; (ii) for difficult cases (i.e., low-resolution images), the classification at the classes with the finest details may fail, but the system could provide robust results for more abstract classes (upper levels in the hierarchy).
\end{description}

In this paper, we experimentally analyze the impact of these components when training a CNN for morphology-based galaxy classification.
To perform this study, we used the Galaxy Zoo 1 \citep{Lintott:08} dataset as a testbed, which provides an excellent laboratory to illustrate our general results. These can be found in 
more complex classification schemes and other sky surveys, so our analysis will be potentially useful for future works that train deep learning models from crowd sourcing.

This paper is structured as follows. Section \ref{rw} summarizes previous work on galaxy classification, with emphasis on deep learning techniques. Section \ref{disagr} gives an introduction to learning from disagreement; Section \ref{gz} introduces Galaxy Zoo 1. The methodology is presented in Section \ref{method}, followed by the experimental results in Section \ref{results}. Section \ref{disc} presents the main lessons obtained from the experiments. A summary and some directions for future work are given in Section \ref{concl}. Appendix \ref{app-res} provides several tables that give more details of the results, and Appendix \ref{sec:si} includes some sample images in which the crowd and the CNN-based classifier differ.

\section{Related work}\label{rw}

Automatic morphological galaxy classification started in the 1990's with  traditional machine learning approaches based on manually engineering features \citep[e.g.,][]{Spiekermann:92} and evolved progressively \citep[e.g.,][]{Huertas:08}. The accuracy of these first approaches was moderate, as the manually selected features were not enough to distinguish the different classes of galaxies. In 2015 appeared some works using deep learning (DL), in particular convolutional neural networks \citep{Dieleman:15,Huertas:15}. Because of major improvements, CNNs have become the state of the art for classifying galaxies based on morphology: \cite{Dieleman:15} provide the first catalog of galaxy morphologies in SDSS (Sloan Digital Sky Survey); \cite{Ghosh:20} performed the classification of distant galaxies from the CANDELS survey; \cite{Vega:21} classify images in the Dark Energy Survey; \cite{Bom:21} worked on the S-PLUS survey; and \cite{Walmsley:24} classify galaxies in several surveys, including SDSS, DESI, CANDELS and UKIDSS.
Other architectures beyond CNNs have been tested for morphological image classification, such as ResNets \citep{Kalvankar:20} and Capsule Networks \citep{Katebi:19}; however, their performances are similar  or inferior to that provided by CNNs. 

\cite{Pandya:23} propose the use of Group Convolutional Neural Networks (GCNNs) \citep{Cohen16} to improve the robustness of morphological galaxy classification, in noisy images and under adversarial attacks (when a small change to the image alters the classification results), applied to Galaxy Zoo DECaLS \citep{Walmsley:22}. \cite{Pandya:23} construct GCNNs utilizing the symmetries of the 2D Euclidean group, which contains all rotations, translations, and reflections in flat space for the task of galaxy morphology classification. The experimental results show that the GCNN outperforms the CNN baseline, at the cost of increasing the computational resources for the training of the model. 

A critical issue in DL is to have enough labeled data in place to properly train the models. \cite{Walmsley:20} proposed active learning as a way to reduce the number of examples required; \cite{Ghosh:20} explored transfer learning; \cite{Vega:21} used a simulated training set; \cite{Khramtsov:22} and \cite{Serrano:24} included data augmentation to generate additional training images. A related issue is how to transfer a model learned from one dataset (i.e., one survey) to another if the instrumental configuration changes, since it affects DL performance. \cite{Khramtsov:22} applied  adversarial validation to analyze the homogeneity of two datasets, so the galaxies of one data set (training) are selected that most closely match the other data set (inference).  \cite{Walmsley:24} jointly trained a classifier from several Galaxy Zoo datasets using multitask learning. \cite{Siudek:25} realized a morphological classification of approx. 300,000 galaxies from Euclid Q1 Release taking advantage of a pre-trained multi-modal foundational model.

 Another related issue is image preprocessing for noise reduction. \cite{Mirzoyan:25}, \cite{Mirzoyan:26}, propose artificial intelligence techniques for denoising galaxy images. The first one presents a U-Net Variational Autoencoder (VAE) framework to denoise galaxy images taken from the James Webb Space Telescope. The second one combines a VAE to remove instrumental contaminants with a GCNN classifier to distinguish disk-like galaxies for non-disks. Noise reduction could be applied as a preprocessing step and could improve later classification results; however, this aspect is beyond the scope of our contribution.

To train a supervised DL model, a set of manually labeled images is required. Initially, datasets were labeled by professional astronomers. \cite{Fukugita:07} compiled a dataset of approx. 2500 galaxies from the SDSS survey. \cite{Schawinski:07} identify elliptical galaxies from a sample (5\%) of the SDSS Data Release 4. Having experts label galaxies in the recent sky surveys has become an overwhelming task, so the scientific community has relied on crowd sourcing, inviting large numbers of people to classify galaxies over the internet. Based on this scheme, the Galaxy Zoo project emerged, in which thousands of volunteers have participated labeling galaxy images according to a well stablished protocol. For SDSS, two datasets based on crowd annotations emerged: Galaxy Zoo 1 \citep{Lintott:08}, in which volunteers categorize images into 6 categories; and Galaxy Zoo 2 \citep{Willett:13}, which included a more complex annotation scheme, based on a series of questions with 11 classification tasks and a total of 37 possible responses. Training machine learning models for morphological classification of images from the SDSS and other surveys relies on crowd annotations. However, crowd labeled data can produce errors and this could affect machine learning models trained on these data, as we will explore in the following section.

\section{Learning from disagreement}\label{disagr}

\subsection{Single level agreement}

In contrast to common supervised learning, learning from crowd annotations introduces uncertainty on the labels of the training data. There could be disagreement between the annotations provided by the crowd for different reasons: (i) Annotator errors; (ii) Interface problems; (iii) Imprecision or incompleteness in the annotation scheme; (iv) Ambiguity; and (v) Item difficulty, i.e., computer vision: low resolution; natural language processing: different views.
There are several approaches to learning from crowd annotations \citep{Uma:21}, which can be divided into four main categories:
\begin{enumerate}
\item Methods that aggregate crowd annotations into a single label for each object, for example, using a majority vote. These are known as  silver labels (in contrast to gold labels that are usually given by experts). Their reliability improves as more annotators perform the classification.
\item Methods that use information on disagreement to eliminate items with low agreement or weight them.
\item Methods that learn a classifier directly from crowd annotations via a probability distribution assigning a score to each label. This is known as soft labels, in contrast to the two previous approaches that produce hard labels.
\item Methods that combine hard and soft labels.
\end{enumerate}

\begin{figure}[h!]
   \centering
   \includegraphics[width=\hsize]{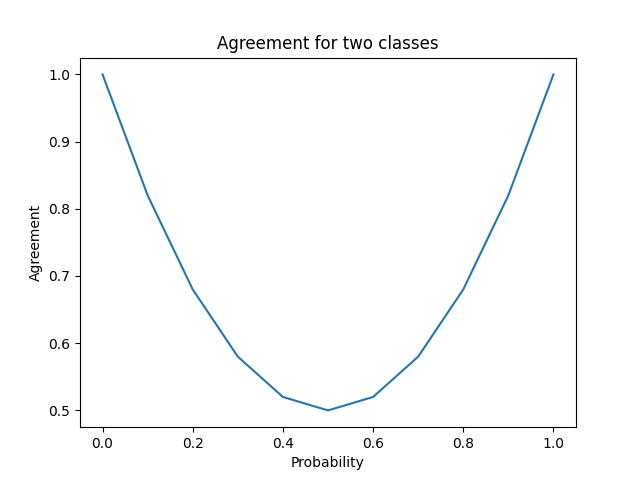}
      \caption{Agreement vs. probability for two classes.}
         \label{agr-prob}
   \end{figure}

Here, we adopt the second approach, for which we need to estimate the agreement for each item.
 \cite{Artstein:08} propose an equation 
 to calculate the observed agreement. Given a set of items $I$ indexed by $i$, a set of categories $K$ indexed by $k$, and a set of annotators $C$ indexed by $c$, the observed agreement for each item, $agr_i$, is given as:
\begin{equation}\label{eq:agrori}
    agr_{i} = \frac{1}{c_i(c_i - 1)} \sum_{k=1}^{K } n_{ik} (n_{ik} - 1),
\end{equation}

\noindent
where $n_{ik}$ is the number of times item $i$ is classified as category $k$.
If we do not have information regarding the number of annotators for each category, we can approximate Eq. \ref{eq:agrori} in terms of the probabilities for each category (estimated from the fraction of annotators that selected the category), with Eq. \ref{eq:agree}:
\begin{equation}\label{eq:agree}
    agr_{i} = \sum_{k=1}^{K}P(k)_{i}^{2},
\end{equation}
where $P(k)_i$ is the probability according to  the percentage of volunteers that classified the $i$-th galaxy with the $k$-th class. This function is the inverse of the entropy, so a higher agreement implies a lower entropy. The agreement for two classes against the probability of one class (the other is simply the complement to one) is illustrated in Figure \ref{agr-prob}.  When one of the classes has a probability of one, $agr = 1$, and when all the classes have equal probabilities, $agr = 1/k$, where $k$ is the number of classes (this will tend to zero when $k$ tends to infinity).

In this work, we  analyze the impact of selecting different levels of agreement between the annotators for training a deep neural network for morphological image classification of GZ1.  We selected agreement thresholds as it is more direct to train machine learning models from hard (silver) labels than soft labels; we also wanted to compare the performance across different agreement ranges. On the other hand, analysis of inter-coder agreement \citep{Dawid:79, Artstein:08} is out the scope of this work since it is required to know the individual votes of each annotator or of pairs of annotators who classified the same set of instances, something that hardly happens when hundreds of thousands of instances are available to be classified.

\subsection{Hierarchical agreement}

The previous measure of agreement, Equation \ref{eq:agree}, applies when all categories are at the same level, as is the case for GZ1, in which the annotations are only for the leaf nodes in the hierarchy. When annotations are provided at different levels in the hierarchy, as is the case for Galaxy Zoo 2 \citep{Willett:13}, another measure of agreement is required that considers partial agreements among dimensions, which we denote as  hierarchical agreement. Next, we propose and analyze two measures of hierarchical agreement as a foundation for future work on more complex classification tasks.

Initially, the Weighted Hierarchical Agreement (WHA) measure is proposed to extend the agreement to hierarchical label spaces. 

\begin{equation}
    \mathrm{WHA} = \sum_{k\in\mathrm{Leaves}} P'(k)^{2}, \label{eq:wha}
\end{equation}

\noindent
where $Leaves$ is the set of leaf nodes in the hierarchy, and $P'(k)$ is the fraction of annotators that classify the instance as class $k$. $P'(k)$ is estimated recursively as:

\begin{equation}
    P'(k) =
    \begin{cases}
        P(k), & \text{if \(x\) has no parents},\\[4pt]
        P(k)\displaystyle\sum_{y\in\mathrm{Parents}(k)} P'(y), & \text{otherwise},
    \end{cases}
    \label{eq:probx}
\end{equation}

\noindent
where $P(k)$ denotes the probability of node $k$ in its corresponding dimension.

Basically, WHA corresponds to the classical agreement (Eq. \ref{eq:agree}) computed over leaf nodes in the hierarchy, weighted by the probabilities of the nodes in the path from the root to each leaf node. This measure is appropriate for tree hierarchies, such as GZ1; but has limitations when applied to directed acyclic graph (DAG) hierarchies such as GZ2, as in this case there are several paths that can reach a leaf node, and WHA does not account for agreement at internal nodes.

Thus, another way of estimating the agreement in a multidimensional scenario, i.e. a scenario with several dimensions (questions) and each one with its own set of classes, is proposed. Multi-Dimensional Weighted Agreement is defined as:

\begin{equation}
    MDWA = \frac{\sum_{i=1}^{D}(w_{i} \cdot agr_{i})}{\sum_{i=1}^{D} w_{i} }, 
\end{equation}

\noindent
where $D$ is the number of dimensions, $agr_i$ is the classical agreement for the {\em i-th} dimension and $w_i$ is the weight (the number or fraction of annotators in the {\em i-th} dimension). That is, MDWA is a combination of the agreements of the different dimensions, which are weighted by the number of annotators in each dimension.
However, considering that each dimension may have a different number of
classes, the agreement may be biased toward dimensions (questions) with fewer number of classes. For instance, suppose two dimensions, the first has 2 classes while the second has 5, the weights are $w_1 = w_2 = 1$ and there is perfect disagreement in both dimensions, that is, $agr_1 = 0.5$ and $agr_2 = 0.2$, therefore, $MDWA = 0.35$ is the agreement score. This example highlights how dimensions with fewer classes may disproportionately influence the overall agreement score, even when disagreement is equally strong across dimensions. This motivates the need for a normalization strategy. The following scale factor is applied to each dimension, so it is normalized to $[0, 1]$:

\begin{equation}
    ScAgr_{i} = \frac{agr_{i}-\frac{1}{|K_{i}|}} {1-\frac{1}{|K_{i}|} }, 
\end{equation}

\noindent
where $agr_i$ is the classical agreement for the {\em i-th} dimension, and $K_i$ is the set of classes of the corresponding dimension. Therefore, the normalized agreement is defined as:

\begin{equation}
    NMDWA= \frac{\sum_{i=1}^{D}(w_{i} \cdot ScAgr_{i})}{\sum_{i=1}^{D} w_{i} } \label{eq:mdwa}.
\end{equation}
\noindent
NMDWA requires that the probabilities in each dimension add up to one.

\begin{figure}[t]
   \centering
   \includegraphics[width=\hsize]{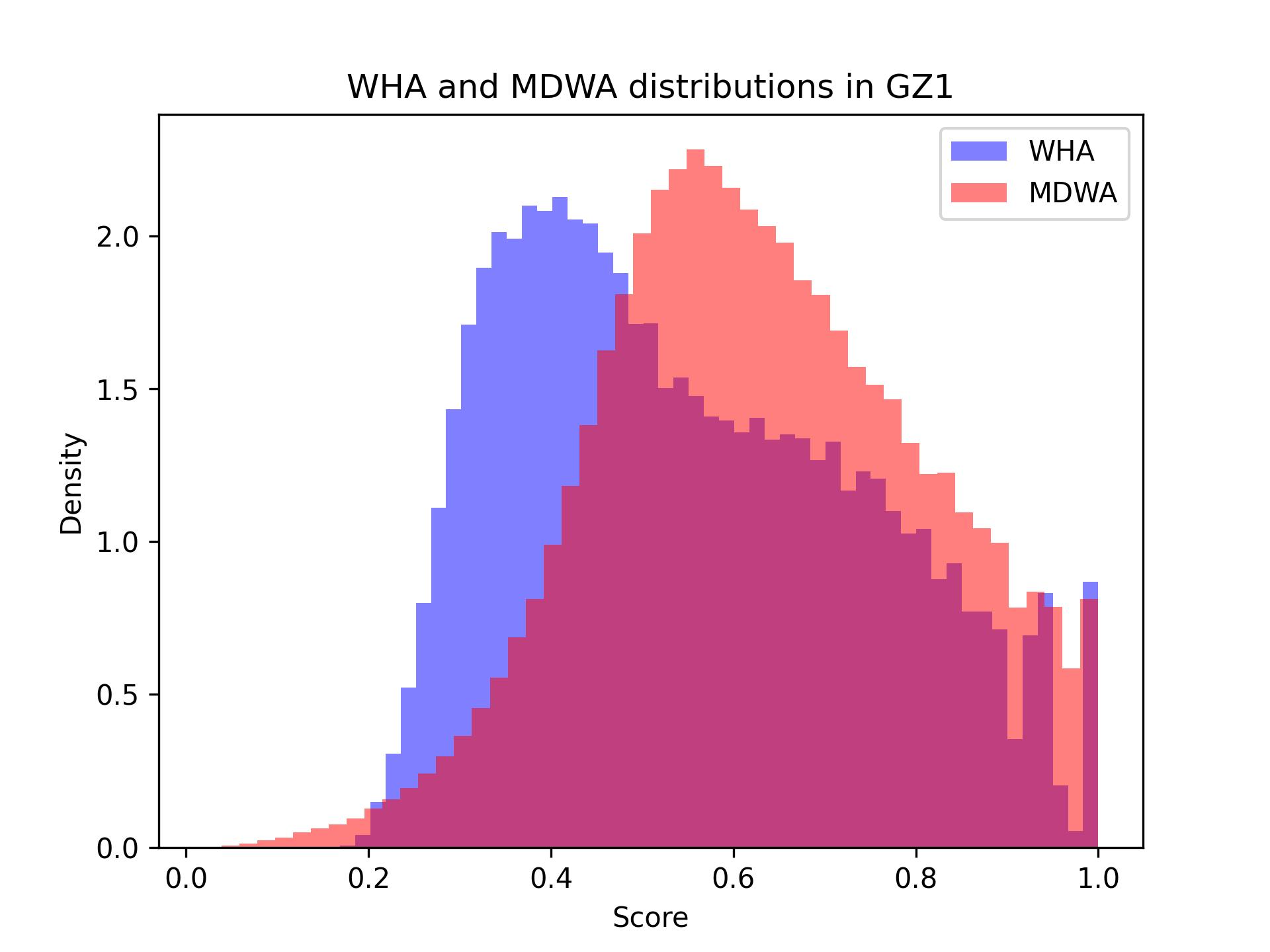}
      \caption{Distribution of agreements computed with WHA and NMDWA for the annotated instances in Galaxy Zoo 1.}
         \label{gz1-ha}
   \end{figure}

Both proposed hierarchical agreement measures, WHA and NMDWA, work with hierarchies of tree type, so we compute both for the Galaxy Zoo 1 hierarchy, illustrated in Figure \ref{hierarchy}. WHA and NMDWA were computed for all instances in the Galaxy Zoo 1 annotated data set (see section \ref{gz}); Figure \ref{gz1-ha} shows the distribution of WHA and NMDWA.

While both metrics yield identical counts for perfect agreement ($agr = 1$), NMDWA consistently assigns higher values than WHA in the intermediate agreement ranges, particularly between $(1, 0.5]$. This trend suggests that NMDWA places greater emphasis on moderate agreements, amplifying their contribution relative to WHA. Conversely, in lower agreement intervals such as $(0.5, 0.3]$, NMDWA values drop sharply, indicating a more selective weighting scheme. The Pearson correlation coefficient between MDWA and WHA was found to be 0.9, indicating a strong positive linear relationship between the two metrics. This high correlation suggests that, despite their different weighting schemes, both measures tend to rank agreement levels similarly for tree-hierarchies. However, as mentioned before, in the case of DAG-hierarchies, NMDWA is required.

\section{Galaxy Zoo 1}\label{gz}

\begin{figure}[t]
   \centering
   \includegraphics[width=\hsize]{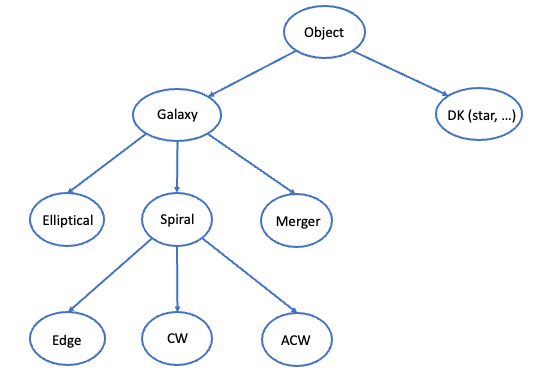}
      \caption{Galaxy hierarchy considered in the experiments for GZ1.}
         \label{hierarchy}
   \end{figure}

Galaxy Zoo 1 (GZ1) \citep{Lintott:08} was drawn from the SDSS galaxy survey, which includes 893,212 objects, of which 667,944 are in the crowd-annotated dataset. In the annotation process, nearly 100,000 annotators provided 38 annotations per image on average. Six classes were considered by the annotators: (i) elliptical, (ii) clockwise spiral (CW), (iii) anticlockwise spiral (ACW), (iv) other (e.g. edge-on, Edge), (v) star or don't know (DK), (vi) merger. These classes can be comprehended by the hierarchy shown in Figure \ref{hierarchy}.

Once the annotation campaign was completed, certain post-processing was performed by the Galaxy Zoo team. First, they estimated the reliability of the annotators, giving more weight to those who provided more consistent annotations (agreement with the majority).
Another issue is that small, faint, or distant galaxies are likely to appear as elliptical and therefore could be classified as such, but these
are likely to be spirals whose arms are not distinctly visible. By assuming that the morphological fraction within bins of fixed galaxy size and luminosity does not evolve over the depth of the survey, it is possible to statistically estimate the bias affecting the morphological classifications for galaxies of known luminosity, size, and distance. This bias is corrected based on the redshift obtained from the spectroscopy data, so the proportion of each class is approximately the same for different redshifts \citep{Lintott:11}.
In our experiments, we used the original fractions of votes for each class, as the bias-corrected ones were not available for all classes.

The galaxy images were downloaded from SSDS DR7\footnote{Although the crowd annotations were realized on images from DR6, this release was no longer available when these experiments were performed.}. 

\section{Methodology}\label{method}

Galaxy classification was performed in two stages.
First, pre-trained CNN {\sc EfficientNetV2} \citep{Tan:21} was adopted. Such a CNN is a type of feedforward neural network that learns features via filter optimization. This type of neural networks is particularly useful for processing vector or matrix data, such as text, images, and videos; and have become de-facto standard in deep learning-based approaches to computer vision and image processing. {\sc EfficientNetV2} is a family of convolutional networks that have faster
training speed and better parameter efficiency than previous models, and at the same time show high accuracy in several image classification tasks.

{\sc EfficientNetV2} (XL) was used in the experiments; it has 208 million parameters and was originally trained with ImageNet data \citep{Deng:09}.
The model was modified for galaxy classification with one output per class and provides an estimate of the probability for each class. Two training schemes were considered: (i)  Last-layer training (hereafter {\em Last}): the weights of the pre-trained CNN were frozen, so only the weights of the last layer were updated. 
(ii) All-layer training (hereafter {\em All}): the weights of all the layers of the pre-trained CNN are updated. 
For both training schemes, the set of images was split into training, validation and test sets, corresponding to 64\%, 16\% and 20\%, respectively. 

In a second stage, the initial classification obtained by the CNN is refined considering the hierarchy.
 Hierarchical classification is a type of multidimensional classification (outputs a set of classes instead of a single class) in which the classes are structured in a hierarchy, a tree or a directed acyclic graph \citep[see the review of][]{Silla:11}. By considering the hierarchy, classification performance can be improved, so in a second stage, a Bayesian network representing the hierarchy is used \citep{Serrano:24}. A Bayesian network is a graphical model that compactly represents the joint probability distribution of a set of random variables, by codifying the conditional independence relations between variables through a directed acyclic graph or DAG \citep{Sucar:21}. The Bayesian network  combines the probabilities obtained from each class via probability propagation, guaranteeing that the final probabilities are consistently inferred. 
Thus, the class of a variable is consistent with the class of all its ancestors in the hierarchy. The classification model that integrates both stages is called  Bayesian Convolutional Neural Network (BCNN). See \citet{Serrano:24} for more details. 

Once the probabilities of all the classes in the hierarchy are updated by the Bayesian network, a  path in the hierarchy (from the root to a leaf) is returned as a result for each image using a top-down approach. That is, in each level, the class with highest probability is selected until a specific class (leaf) is reached. The classifier returns a set of classes that satisfy the hierarchical constraint. In the experiments, we compare the results of using only the CNN or incorporating the hierarchy with the BCNN.

To evaluate the impact of quality vs. quantity in the classification accuracy, we train the classifier with different configurations of training data based on the crowd agreement ({\em agr}). Higher agreement implies a higher quality of the training data, as more annotators agree on the class; however, it also implies a reduction in the amount of training data, which has an impact on deep learning techniques. This also depends on the intrinsic difficulty of the classification problem and the bias introduced in the annotation process.  We consider different thresholds on the agreement: 1 (100\% agreement), 0.90, 0.80, ..., 0.3. For evaluation we consider the test images for the corresponding range,  
and report the hierarchical evaluation measures.

Additionally, we evaluate a transfer learning approach, in which the model is trained based on exclusive ranges; that is, for $agr =1$, then for $1 > agr \geq 0.9$, $0.9 > agr \geq 0.8$, etc., until $0.4 > agr \geq 0.3$. The model trained in each range is used as an initial one for the next range, starting from  $agr =1$. This is known as  curriculum learning \citep{Soviany:22}, as the network is trained first with the  easy cases, and then more  challenging cases are incorporated. We also test the impact of a reduced training set (200 to 1000 training data), with both approaches, CNN and BCNN.

Finally, we also consider combining several models into an  ensamble of classifiers. A test image is given simultaneously to $N$ pre-trained models, and the outputs are combined to produce the final classification. There are several combination schemes, we use a simple majority voting, considering 3 and 5 models.

\section{Experimental results}\label{results}

\subsection{Evaluation measures}

In this work, five evaluation measures commonly used in hierarchical classification are employed. The  exact match (EM) is the most strict evaluation measure because the prediction of all the classes in the path in the hierarchy must be equal to the set of real labels (all classes are correct on the path). However, in hierarchical classification, the predictions may be partially correct; in this way, hierarchical accuracy,  precision, recall, and F-measure give a score (greater than zero) if the prediction is partially correct. For example, the classifier could be wrong regarding the most specific class in the hierarchy (i.e. ACW), but correct regarding higher level classes (Spiral, Galaxy), and these other measures take these partially correct answers into account.

Let $N$ be the number of instances in the test set, $Y$ be the real subset of classes with which an instance is associated, and let $\hat{Y}$ be the subset of predicted classes. 
The evaluation measures are defined as follows.\\
$Exact\ match$:
\begin{equation}
EM = \frac{1}{N} \sum_{i=1}^N 1_{Y_i = \hat{Y}_i}.
\end{equation}
\noindent
$Hierarchical\ accuracy$:
\begin{equation}
hA = \frac{1}{N} \sum_{i=1}^N \frac{ | Y_i \cap \hat{Y}_i |}{| Y_i \cup \hat{Y}_i |}.
\end{equation}
\noindent
$Hierarchical\ precision$:
\begin{equation}
hP = \frac{\sum_{i=1}^N  | Y_i \cap \hat{Y}_i |}{\sum_{i=1}^N  | \hat{Y}_i |}.
\end{equation}
\noindent
$Hierarchical\ recall$:
\begin{equation}
hR = \frac{\sum_{i=1}^N  | Y_i \cap \hat{Y}_i |}{\sum_{i=1}^N  | Y_i |}.
\end{equation}
\noindent
$Hierarchical\ F-measure$:
\begin{equation}
hF = \frac{2 * hP * hR}{hP + hR}.
\end{equation}

In our experiments, the real subset of classes to which an instance is associated corresponds to the path that ends in the leaf node that received the majority of votes as the community correct (silver) label \citep{Uma:21}.

\subsection{Results and analysis}

\subsubsection{Training on cumulative ranges of agreement}

In this first experiment, we train the model with several cumulative ranges of agreement, starting from 100\% agreement ($agr = 1$), then applying $agr \geq 0.9$, $agr \geq 0.8$... $agr \geq 0.3$. We evaluate the models in the test sets for the corresponding agreement ranges with the different hierarchical evaluation measures. We present the results for the different agreements and quantity of training images, considering two additional dimensions: (i) if only the CNN is used or if the  BCNN is used; (ii) training only the last layer of the CNN ({\em Last}) or training all the layers ({\em All}). 

\begin{figure}[h!]
   \centering
   \includegraphics[width=\hsize]{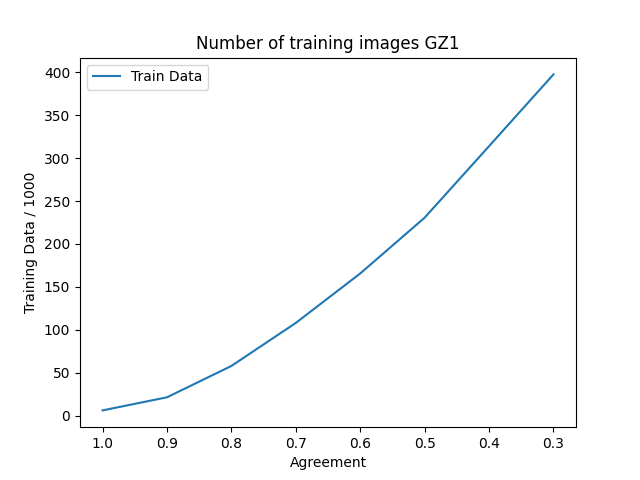}
      \caption{Number of training images for each cumulative range of agreement.}
         \label{traindata-acc}
   \end{figure}

\begin{figure}[h!]
   \centering
   \includegraphics[width=\hsize]{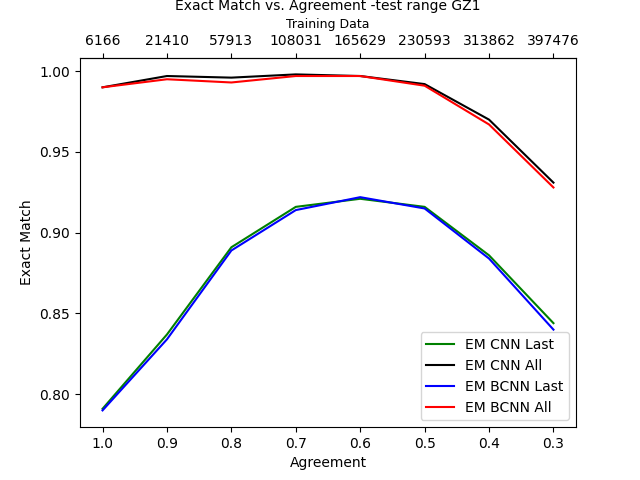}
      \caption{Results in terms of  Exact Match for each of the different levels of cumulative agreement and quantity of training data for the CNN and BCNN classifiers, retraining the last layer ({\em Last}) or all the layers ({\em All}).}
         \label{EM-testrange-acc}
   \end{figure}

Figure \ref{traindata-acc} shows the number of training data per cumulative range of agreement (CRA). Figure \ref{EM-testrange-acc} shows the exact match vs. CRA and training data, considering CRA for different thresholds from $1$ to $0.3$. 
This figure shows a peak $EM \approx 0.92$ for $CRA=0.6$ if re-training the last layer, while the peak is significantly higher (EM $\approx 1.00$) and achieved with less data but higher agreement ($CRA=0.7$) if re-training all layers, both for CNN and BCNN.
Appendix \ref{app-res} provides additional information: (i) Table \ref{t:ds_class} with the number of training, validation, and test images per class, (ii) Tables \ref{res-ll-range} and \ref{res-al-range} with the results for the different evaluation measures for the cumulative test sets.

{\bf Analysis}:
There is a high performance (exact match above 99\%) when re-training the full network with agreements above $0.6$, and it starts to decrease with lower agreements. We notice a different behavior when only re-training the last layer. In the last scenario, the exact match increases until $CRA = 0.6$, and then it starts to decrease. This could be because there is a significant amount of more training data, and this is important when only the last layer is re-trained.

From the results, we observe an important difference between retraining only the last layer vs. retraining all layers in the neural network,
with a significant improvement of about 10\% points for all measures. As indicated by \citet{Walmsley:24}, this could be due to differences between the characteristics of the galaxies and those of the images used originally to train the neural network, so the learned traits are not optimal for galaxy classification.

The results incorporating the hierarchy in a post-processing stage (BCNN), are similar to those obtained without it (CNN). This could be because there are few classes and the hierarchy is  small; and also as we increase the amount of training data the performance of the CNN increases and the impact of the BCNN decreases. The hierarchical classifier could be beneficial when the amount of data is reduced, as we will show later. 
Also, the hierarchical measures show higher performance, implying that although the classification is not perfect, it is partially correct in some levels of the hierarchy, providing information about the type of galaxy at a higher level of abstraction. See Appendix \ref{app-res}.

\subsubsection{Training on exclusive ranges of agreement}

\begin{figure}[h!]
   \centering
   \includegraphics[width=\hsize]{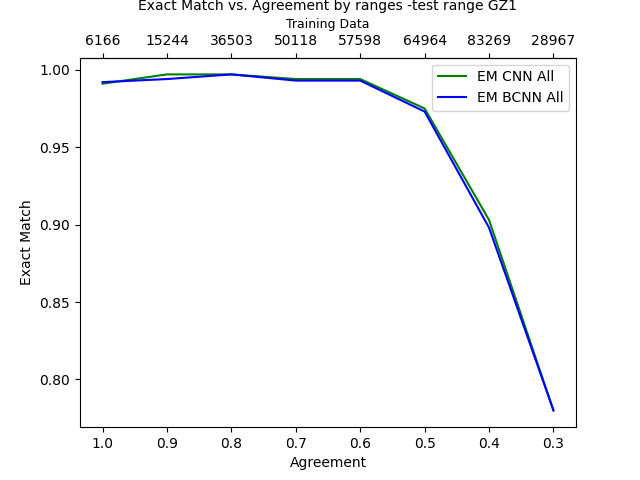}
      \caption{Results in terms of Exact Match for ERA considering the different ranges of agreement and quantity of training data, for the CNN and BCNN classifiers retraining all the layers.}
         \label{EM-ranges}
   \end{figure}

In this second experiment, we train the model from the images in the exclusive ranges of agreement (ERA); that is, for $agr =1$, then for $1 > agr \geq 0.9$, $0.9 > agr \geq 0.8$, and so on, until $0.4 > agr \geq 0.3$. In contrast with the previous experiment, in this one we apply  transfer learning, so the model trained for each range is used as the initial model for the following range. In this case, we only consider updating all the layers of the neural network. The results are depicted in the following figures. Figure \ref{traindata-range} shows the number of training data per range of agreement. Figure  \ref{EM-ranges} shows the exact match vs. the agreement and training data, considering the different ERA for the CNN and BCNN classifiers.

\begin{figure}[t!]
   \centering
   \includegraphics[width=\hsize]{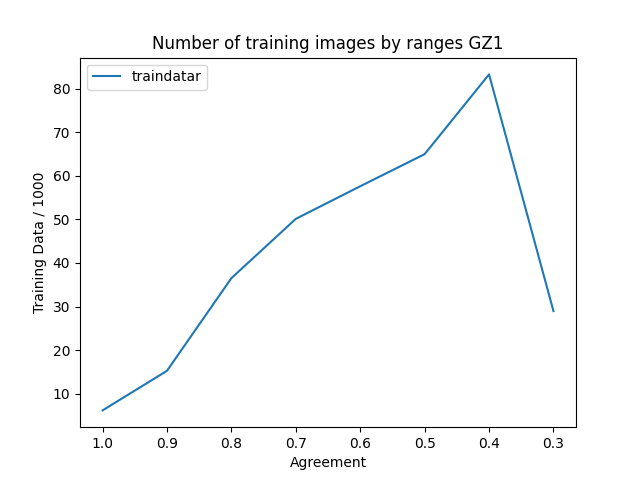}
      \caption{Number of training images for each exclusive range of agreement.
      }
         \label{traindata-range}
   \end{figure}

Appendix \ref{app-res} provides additional information: (i) Table \ref{num-ranges} with the number of training, validation, and test images per agreement range, (ii) Table \ref{res-ranges} with the results for the different evaluation measures for the ERA test sets.

\begin{figure}[t!]
   \centering
   \includegraphics[width=\hsize]{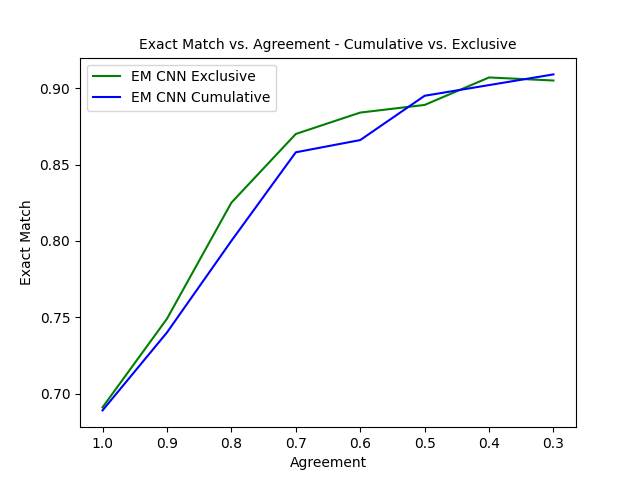}
      \caption{A comparison of training in an cumulative scheme vs. training by ranges with transfer learning, when the evaluation is done with all the test images.}
         \label{EM-comp-testall}
   \end{figure}

{\bf Analysis:} We notice a similar performance training with exclusive ranges of agreement as when training in a cumulative scheme; however, to make a fair comparison, we need to compare the models in the same set of test data. Thus,  we consider the evaluation on the full test sets (presented below, see section \ref{sec:fulltest}). For ease of comparison, we show both graphs together in Figure \ref{EM-comp-testall}. 
We notice a similar performance, although a small advantage for some cases when training by ranges with transfer learning, in particular for less than $200,000$ training data. This could be because starting from the  easy cases (high agreement) to the more difficult ones,  can provide better performance (curriculum learning). With more data, both training schemes converge to approx. the same results.

\subsubsection{Training on reduced data}

\begin{figure}[b!]
   \centering
   \includegraphics[width=\hsize]{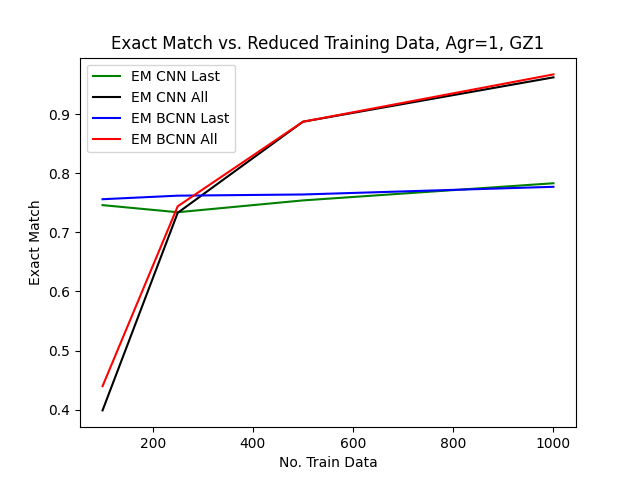}
      \caption{Results in terms of Exact Match for a reduced number of training images with $agr =1$ for the CNN and BCNN classifiers.}
         \label{EM-red-data}
   \end{figure}

In this experiment, we analyze the impact of a significant reduction of the number of images in the training data set.
We consider the set of test images with perfect agreement, and choose random subsets of training images, from 1000 to 100 images.
Results are depicted in Figure \ref{EM-red-data},  considering two additional dimensions: (i) if only the CNN is used or if the BCNN is used; (ii) training only the last layer of the CNN (Last) or training all the layers (All). 

{\bf Analysis:} For a relatively large amount of training data, above 5,000 samples, and $agr \geq  0.6$, the results are very good when retraining the entire network. Reducing the training data still obtains good results up to 1000, although the evaluation measures start to decrease. For less data, there is a more significant decrease in performance, although still relatively good, above $75$\%, with only 100 images, less than 20 per class in average.

We can also notice that with reduced training data, there is an impact of incorporating the hierarchy with the BCNN classifier, showing higher exact match than the CNN classifier, in particular when retraining only the last layer. This could be because the effectiveness of CNN is highly dependent on the amount of training data and the information provided by the hierarchy helps to reduce the impact of having few training data.

In contrast to when the number of training data is  large (in this case greater than 250 samples), if the number of training images is small, it is better to train only the last layer, as illustrated in Figure \ref{EM-red-data}. This finding suggests that it is precise to have  enough data to retrain the whole network. Figure \ref{EM-red-data} also indicates that the models in which only the last layer is trained show equal performance regardless of increasing the data used for the training. This suggests that there is no improvement in learning for this limited amount of training data; however, with more data the performance increases, see Figure \ref{EM-testrange-acc}.

\subsubsection{Evaluating on the full test set}\label{sec:fulltest}

\begin{figure}[t!]
   \centering
   \includegraphics[width=\hsize]{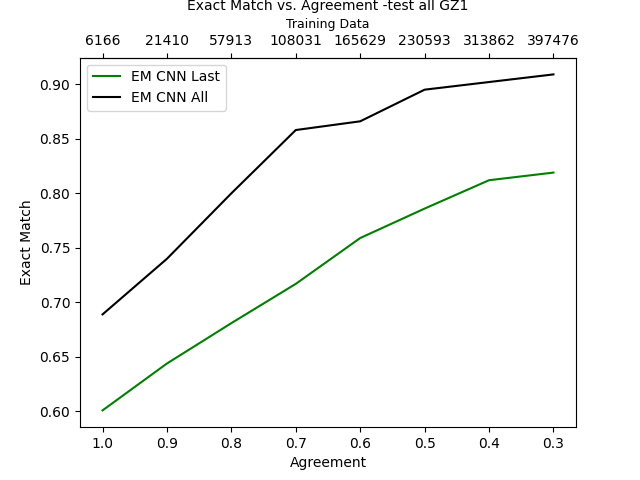}
      \caption{Results in terms of  Exact Match for each of the different levels of agreement and quantity of training data when training is done in a cumulative scheme and the evaluation is done with all the test images, CNN retrained in the last or all layers.}
         \label{EM-testall-acc}
   \end{figure}

\begin{figure}[h!]
   \centering
   \includegraphics[width=\hsize]{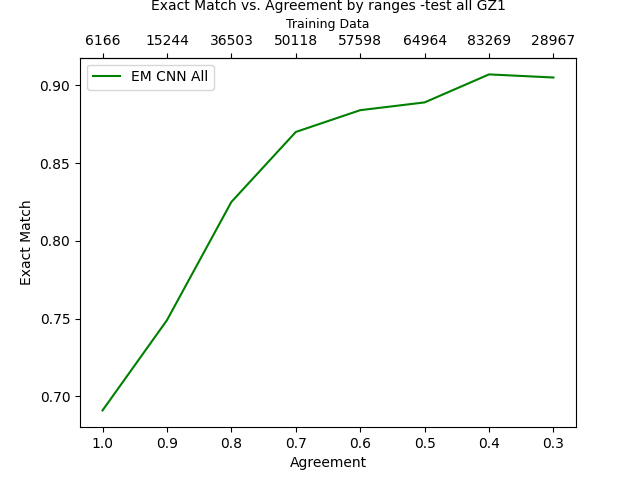}
      \caption{Results in terms of  Exact Match for the different ranges of agreement  and quantity of training data when training is done by exclusive ranges with transfer learning and the evaluation is done with all the test images; the weights of all layers in the CNN are retrained.}
         \label{EM-ranges-testall}
   \end{figure}

As a final experiment, we evaluate the models trained on the different levels of agreement on the full test set, which implies a more challenging scenario. In this case, each model trained on the different levels of agreement (cumulative or exclusive), is evaluated on the full test set (obtained by the union of the test images for the exclusive ranges of agreement in Table \ref{num-ranges}). 
Figure \ref{EM-testall-acc} depicts the exact match for the CRA, while Figure  \ref{EM-ranges-testall} for the ERA.

Appendix \ref{app-res} provides the results for the different evaluation measures. 
Tables \ref{res-cum-all} and \ref{res-al-all} for the models trained on cumulative ranges, and  Table \ref{res-ranges-all}  for the models trained with exclusive ranges.

{\bf Analysis:} Deep learning models depend on having a large and representative set of training samples; thus, when evaluating on the full data set, the model trained on all the galaxies for the different agreement levels achieves the higher accuracy. It obtains an exact match over 90\% including some difficult cases (agreements below 0.4).

\subsubsection{Galaxy type differences}

To compare the performance for the different classes in GZ1, we analyzed the confusion of the classifier between the different classes for the different levels of agreement using a  confusion matrix, in which the number of items predicted for each type of galaxy are compared to the actual number of items for each type (in this case, the class selected by the majority of the annotators). We show the confusion matrices for some of the experiments, in particular for the cases with perfect agreement ($agr = 1$), Table \ref{cm1}; and for $agr \geq 0.7$, Table \ref{cm2} (which has the higher exact match of $99.75$), for both, the CNN and BCNN classifiers retrained in all layers. We notice a significant majority in the main diagonal that indicates a correct classification, with few cases of incorrect classifications. For most classes there is no clear pattern, except for ACW that tends to be confused with MERGER. Proportionally we observe a larger uncertainty for MERGER, as this is the class with less training samples.

\begin{table}[h!]
\centering
\caption{Confusion matrices for $agr = 1$, retraining all the layers}
\label{cm1}
\begin{tabular}{c}
\includegraphics[width=8cm]{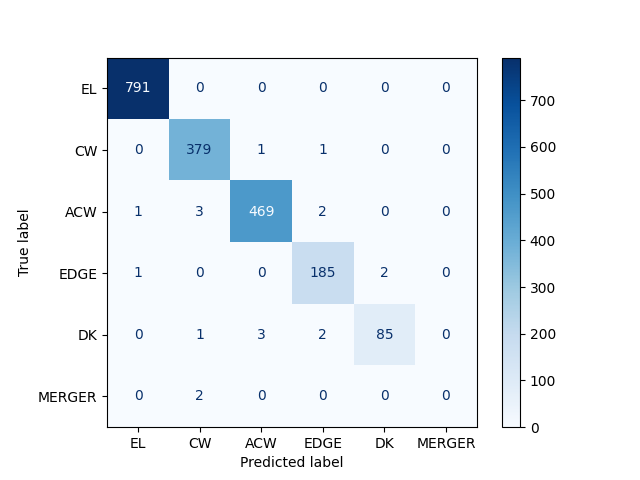} \\
CNN \\ 
\includegraphics[width=8cm]{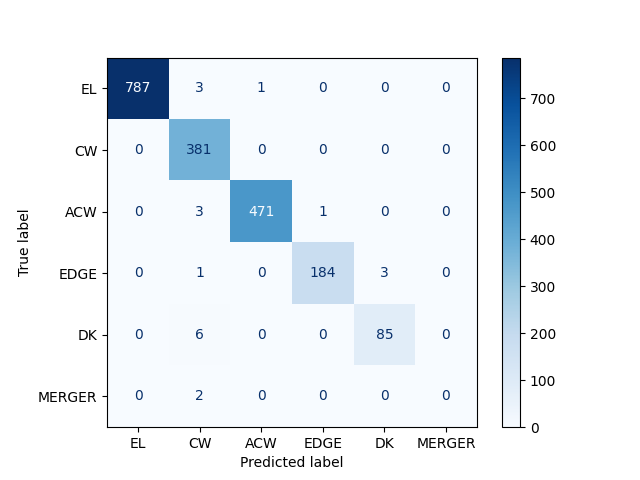} \\
BCNN \\ 
\end{tabular}
\end{table}

\begin{table}[h!]
\centering
\caption{Confusion matrices for $agr \geq 0.7$ with CRA, retraining all the layers}
\label{cm2}
\begin{tabular}{c}
\includegraphics[width=8cm]{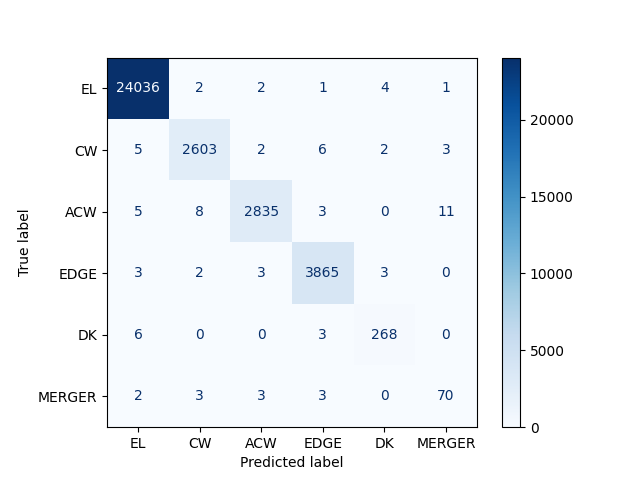} \\
CNN \\
\includegraphics[width=8cm]{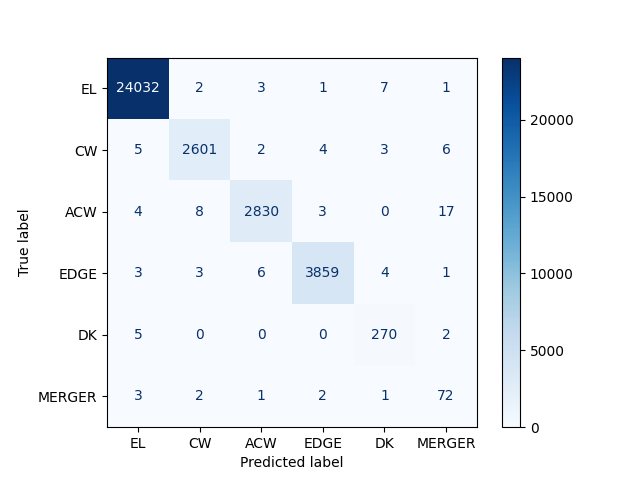} \\ 
BCNN \\ 
\end{tabular}
\end{table}

\subsubsection{Results for ensemble of classifiers}

The five models with the best performance in the full test set are selected to form an ensemble of classifiers. The best models are listed in Table \ref{best-test-all}. The best three and five models are combined using a simple voting scheme to classify the entire set of tests.
Results are shown in Table \ref{t:res_ensemble}.

\begin{table}[h!]
\centering
\caption{Best models in terms of exact match evaluated in the complete test set}
\label{best-test-all}
\begin{tabular}{ lccl } \hline \hline
Model  & Training  & Exact  & Training  \\ 
 Type &  Range &  Match &  Scheme \\ \hline
  CNN  &  $\geq$ 0.3 & 90.91 & cumulative\\
  CNN & (0.5,0.4] & 90.72 & ranges\\
  BCNN & $\geq$ 0.3 & 90.51 &cumulative \\
  CNN & (0.4,0.3] &90.51 & ranges \\
   CNN & $\geq$ 0.4 & 90.21 & cumulative \\ \hline
\end{tabular}
\end{table}

\begin{table}[h!]
\centering
\caption{Results of the ensemble classifier form by the 3 and 5 best models}
\label{t:res_ensemble}
\begin{tabular}{@{}lccccc@{}} \hline \hline
\textbf{Ensemble} & \textbf{EM} & \textbf{Accuracy} & \textbf{hR} & \textbf{hP} & \textbf{hF} \\ \hline
3 best            & 90.86       & 93.45             & 94.59       & 94.31       & 94.45       \\
5 best             & 91.04       & 93.64             & 94.84       & 94.46       & 94.65       \\ \hline
\end{tabular}
\end{table}

{\bf Analysis:} We evaluated combining several models in the more difficult case that uses all the test data.
The results show a small improvement when combining the best five models, against the best individual models.

\subsection{Training Times}

The system is trained on a machine at TeideHPC with the following characteristics: 2 X Intel Xeon Gold 6338 32C” (two 32-core processors each), 256 GB of RAM and an NVIDIA A100 40 GB graphic card.

   \begin{figure}[h!]
   \centering
   \includegraphics[width=\hsize]{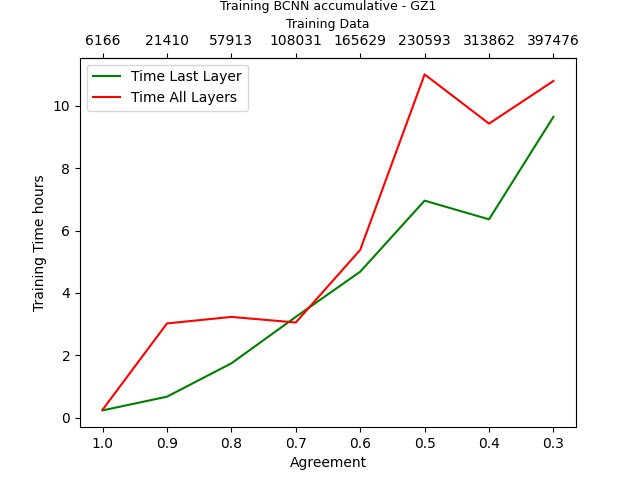}
      \caption{Training time (in hours), for each range of agreement and quantity of training images for cumulative ranges.}
         \label{traintime-cumul}
   \end{figure}

\begin{figure}[h!]
   \centering
   \includegraphics[width=\hsize]{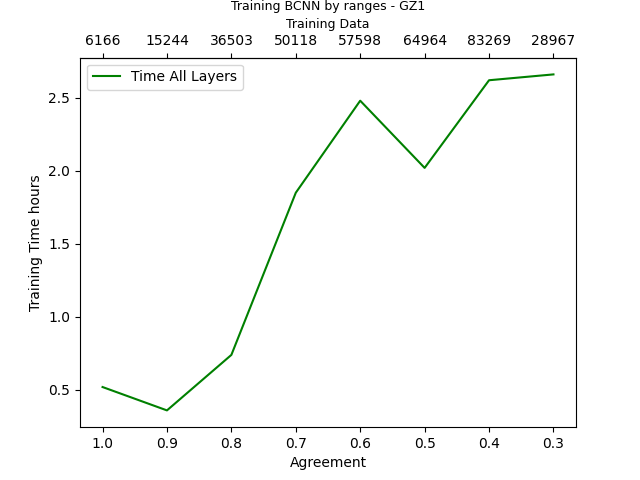}
      \caption{Training time (in hours), for each range of agreement and quantity of training images for exclusive ranges.}
         \label{traintime-ranges}
   \end{figure}

To optimize training times, the models are trained until convergence and not a fixed number of epochs. That is, if the performance does not improve for three consecutive training epochs the training of the model stopped. This implies certain variations in training times that do not increase linearly with the number of training data.  Figure \ref{traintime-cumul} shows the training times per level of agreement and the amount of training data, for the cumulative ranges;  Figure \ref{traintime-ranges} for the exclusive ranges. Although retraining all layers takes more time per epoch, it generally takes less epochs to converge than the model retrained only in the last layer, so the differences between both schemes are not significant. This could be because the model retrained completely provides more flexibility, in contrast to the one in which only the weights in last layer are modified.

\subsection{When the crowd and ML differ}

In Appendix \ref{sec:si} we show some examples of images for which the classification given by the majority of annotators and the CNN differs, including cases with perfect agreement, $agr=1$, and difficult cases, $agr \leq 0.4$. 

For the cases with $agr = 1$, in three of them the automatic system is partially correct but confuses ACW with CW or ACW with EDGE; in the other one, it seems it is confused by a bright star close to the galaxy. 
In Figure \ref{img1} the crowd is correct. The spiral arms are clear in the left, center-right, and right cases. Differential rotation yields the "S"-shaped pattern observed for the ACW case. Regarding the edge-on spiral galaxy, it is clear too; an elliptical galaxy could not be so flat, while a galaxy with a small bulge and an edge-on disk can reproduce the image observed.
The ML system is not perfect, although for the images with perfect agreement from the crowd, it only misses 12 of nearly 2000 test images. An alternative for these cases, will be to train and combine several models (see the Discussion below).

In Figure \ref{img2} there is not an obvious case. The top-left could be elliptical, but not an edge galaxy. Bottom-left could be a spiral, but not clearly edge-on. Bottom, center-right, could be a face-on spiral or a elliptical. All of these cases are certainly difficult. Reasons could be: 1) Lower signal-to-noise ratio; 2) Relatively faint, small, compact galaxies. 
These objects, if they have spectra, could be good examples to take our analysis to the next level: ML applied to photometry + spectroscopy.

\section{Discussion}\label{disc}

Since 2015, several 
studies have applied convolutional neural networks for morphological galaxy classification with fruitful results \citep[see][for a review]{Huertas:23}. 
Rather than comparing our results with any previous work, as the datasets are different, our objective is to explore for the first time the impact of different factors on morphological galaxy classification.  

The first aspect investigated 
is the agreement between annotators, which is relevant to assess 
the reliability of the classification provided by the crowd as well as 
the accuracy of the model trained on these data. From the experimental results for GZ1, we notice that for $agr \geq 0.6$ the EM is above $99 \%$, and there is a rapid decline for lower levels of agreement. In the case of perfect agreement, the model differs from the crowd in only 12 of 1837 test images (i.e. EM of $99.35\%$). This measure of agreement can also serve as an indicator of the utility of the data labeled by the crowd; classifications with low agreement are not very useful, as shown in the example images with $agr \leq 0.4$ in Appendix \ref{sec:si}. We introduce the hierarchical agreement as a basis for future work on GZ2, including annotations at different levels in the hierarchy.

A second aspect is how to train the convolutional neural network. There are basically three options: (i) starting from a pre-trained model, only retraining the last layer, (ii) retrain all the layers in the network, or (iii) train the model from scratch. Our experimental results show that with  sufficient training data it is better to retrain all the network, with significant difference to only retrain the last layer (10\% difference in EM). This makes sense as the pre-trained models are trained on different kind of images and the learned features in the first layers of the neural network are not optimal for galaxy classification. It also confirms the result of a recent work by \citet{Walmsley:24}. Training the network from scratch was not considered, as this will require an even larger labeled data set and more training time. 

Since the pioneering work of \citet{Hubble:26}, galaxy's morphologies have been represented as a hierarchy of classes, but very few works have taken advantage of the hierarchy in the classification model. An exception is \citet{Serrano:24}, which includes a limited hierarchy with a small dataset. In this work, we analyze the impact of including the hierarchy in the classification process and present the results in terms of hierarchical evaluation measures. Although there is no significant improvement in the case of GZ1 by incorporating the hierarchy, this could be because it is simple (3 levels and 8 classes), in contrast to more complex classification schemes such as GZ2. The Bayesian hierarchical post-processing helps to improve the performance of the model when the amount of data is limited, which implies that the hierarchy can be especially useful in these scenarios. A higher impact can be expected on larger and more complex hierarchies such as GZ2. Previous work \citep{Serrano:24} on a larger hierarchy, 4 levels and 10 classes, presents a higher impact of integrating the BCNN. 

A technique that has not been taken advantage of in previous studies on galaxy classification is curriculum learning, in which a model is trained in several stages, starting from easy cases to more difficult ones. This strategy has been widely used in the fields of computer vision, natural language processing, and reinforcement learning; it can effectively solve  non-convex optimization problems and improve the generalization ability and convergence speed of models \citep{Liu:23}. In our experiments, curriculum learning shows a certain advantage over training in an accumulative scheme, particularly for limited data. For instance, curriculum learning achieves an EM of $0.83$ vs $0.80$ for $agr \geq 0.8$ when evaluation is done with all the test images.

Finally, we analyze the combination of multiple classifiers, commonly known as a classifier ensemble, a common approach in machine learning, which has demonstrated the ability to improve classification accuracy in many application domains \citep{Jurek:13}. We evaluated an ensemble of five models for the most challenging scenario, classifying the entire test set including images with very low agreement; the results show an improvement in all evaluation measures with respect to all individual classifiers, obtaining an $EM$ of $91.04\%$, an $Accuracy$ of  $93.64\%$ and a $hF$ of $94.65\%$.

\section{Conclusions}\label{concl}

We have introduced and analyzed several issues that impact  CNN models learned from crowd annotations for morphological galaxy classification. The main lessons learned from the experiments performed in this paper are the following:
\begin{enumerate}
\item There is a significant difference when retraining all layers vs. retraining only the last layer. This is likely more important if the model has been pre-trained with a different kind of images. Although with very limited data, it could be better to only retrain the last layer.
\item Incorporating the hierarchy in a post-processing phase may improve the accuracy for limited amounts of data.
\item Higher levels of agreement between annotators produce better models, 
proven for agreements equal or greater than $0.6$.
\item In the case of GZ1 with 6 classes, 5000 or more training samples provide enough data when retraining the entire model, the accuracy decreases with 1000 or fewer data (for $agr=1$). Of course, this quantity will vary depending on the quantity and complexity of the morphological classes considered.
\item Training by ranges of agreement, from easy to difficult cases as in curriculum learning, can provide some advantages vs. training with all the samples at once.
\item A combination (ensemble) of models shows a small improvement vs. the best model.
\end{enumerate}

Our findings could be useful when training deep learning models from crowd annotations for next coming large sky surveys, particularly those based on wide-field astronomical observations. Machine learning will be necessary to analyze them, as it becomes impractical to perform visual inspection for all data.

In the next paper in this series, we will present a similar analysis for a more complex classification hierarchy, specifically that used in Galaxy Zoo 2.

\begin{acknowledgements}

      The Instituto Nacional de Astrof\'\i sica, \'Optica y Electrónica (Mexico) is gratefully acknowledged for supporting LES during his sabbatical year at the Instituto de Astrof\'\i sica de Canarias (IAC).
      The authors thank Jos\'e Alfonso L\'opez Aguerri (Astrophysics Department, IAC) for fruitful discussions and Ubay Dorta Guerra (Computer Services,  
      IAC) for his assistance. 
      The authors wish to acknowledge the contribution of Teide High-Performance Computing facilities to the results of this research. TeideHPC facilities are provided by the Instituto Tecnol\'ogico y de Energ\'\i as Renovables (ITER, SA). URL: \url{http://teidehpc.iter.es}. 
      CdB acknowledges support from the Agencia Estatal de Investigaci\'on del Ministerio de Ciencia, Innovaci\'on y Universidades (MCIU/AEI) under grant WEAVE: EXPLORING THE COSMIC ORIGINAL SYMPHONY, FROM STARS TO GALAXY CLUSTERS and the European Regional Development Fund (ERDF) with reference PID2023-153342NB-I00/10.13039/501100011033, as well as from a Beatriz Galindo Senior Fellowship (BG22/00166) from the MICIU. The Universidad de La Laguna (ULL) and the Consejería de Econom\'\i a, Conocimiento y Empleo of the Gobierno de Canarias are also gratefully acknowledged for the support provided to CdB (2024/347). JSP acknowledges support from the Agencia Estatal de Investigaci\'on del Ministerio de Ciencia, Innovaci\'on y Universidades (MCIU/AEI) under grant WEAVE: EXPLORING THE COSMIC ORIGINAL SYMPHONY, FROM STARS TO GALAXY CLUSTERS and the European Regional Development Fund (ERDF) with reference PID2023-153342NB-I00/10.13039/501100011033. 
      Funding for the Sloan Digital Sky Survey V has been provided by the Alfred P. Sloan Foundation, the Heising-Simons Foundation, the National Science Foundation, and the Participating Institutions. SDSS acknowledges support and resources from the Center for High-Performance Computing at the University of Utah. SDSS telescopes are located at Apache Point Observatory, funded by the Astrophysical Research Consortium and operated by New Mexico State University, and at Las Campanas Observatory, operated by the Carnegie Institution for Science. The SDSS web site is \url{www.sdss.org}. SDSS is managed by the Astrophysical Research Consortium for the Participating Institutions of the SDSS Collaboration, including the Carnegie Institution for Science, Chilean National Time Allocation Committee (CNTAC) ratified researchers, Caltech, the Gotham Participation Group, Harvard University, Heidelberg University, The Flatiron Institute, The Johns Hopkins University, L'Ecole polytechnique f\'{e}d\'{e}rale de Lausanne (EPFL), Leibniz-Institut f\"{u}r Astrophysik Potsdam (AIP), Max-Planck-Institut f\"{u}r Astronomie (MPIA Heidelberg), Max-Planck-Institut f\"{u}r Extraterrestrische Physik (MPE), Nanjing University, National Astronomical Observatories of China (NAOC), New Mexico State University, The Ohio State University, Pennsylvania State University, Smithsonian Astrophysical Observatory, Space Telescope Science Institute (STScI), the Stellar Astrophysics Participation Group, Universidad Nacional Aut\'{o}noma de M\'{e}xico, University of Arizona, University of Colorado Boulder, University of Illinois at Urbana-Champaign, University of Toronto, University of Utah, University of Virginia, Yale University, and Yunnan University. 
           This publication uses data generated via the Zooniverse.org platform, development of which is funded by generous support, including a Global Impact Award from Google, and by a grant from the Alfred P. Sloan Foundation.

      \end{acknowledgements}

\bibliographystyle{aa}
\bibliography{references}

\begin{appendix}

\section{Additional information of the data and results}\label{app-res}

This Appendix provides additional information regarding the number of images in each range of agreements, and for the other hierarchical evaluation measures for the different experiments: Section \ref{cum} for the cumulative ranges of agreement, Section \ref{ranges}  for the exclusive ranges of agreement, and Section \ref{full} for the evaluation on the complete (across al agreement ranges) test set.

\subsection{Number of images and other evaluation metrics: cumulative ranges of agreement}\label{cum}

\begin{table}[h!]
\centering
\caption{Number of images associated to each class for each cumulative range. }
\label{t:ds_class}
\resizebox{\columnwidth}{!}{
\begin{tabular}{@{}cccccccccc@{}}
\hline \hline
Agr.          & Set & EL     & CW    & ACW   & EDGE  & DK   & MG & SPIRAL & GX \\ \hline
1  & Tr  & 2532   & 1220  & 1519  & 600   & 289  & 6      & 3339   & 5877     \\
                     & Val & 633    & 305   & 380   & 150   & 73   & 1      & 835    & 1469     \\
                     & Te  & 791    & 381   & 475   & 188   & 91   & 2      & 1044   & 1837     \\ \hline
$\geq$0.9 & Tr  & 11027  & 3401  & 3933  & 2414  & 613  & 22     & 9748   & 20797    \\
                     & Val & 2757   & 850   & 984   & 604   & 154  & 5      & 2438   & 5200     \\
                     & Te  & 3446   & 1063  & 1229  & 755   & 193  & 7      & 3047   & 6500     \\ \hline
$\geq$0.8 & Tr  & 37223  & 6100  & 6774  & 6917  & 790  & 109    & 19791  & 57123    \\
                     & Val & 9307   & 1525  & 1694  & 1729  & 198  & 27     & 4948   & 14282    \\
                     & Te  & 11633  & 1907  & 2116  & 2162  & 249  & 34     & 6185   & 17852    \\ \hline
$\geq$0.7 & Tr  & 76943  & 8386  & 9160  & 12401 & 879  & 262    & 29947  & 107152   \\
                     & Val & 19238  & 2096  & 2291  & 3100  & 220  & 65     & 7487   & 26790    \\
                     & Te  & 24046  & 2621  & 2862  & 3876  & 277  & 81     & 9359   & 33486    \\ \hline
$\geq$0.6 & Tr  & 122649 & 10793 & 11601 & 19034 & 970  & 582    & 41428  & 164659   \\
                     & Val & 30665  & 2698  & 2901  & 4758  & 243  & 145    & 10357  & 41167    \\
                     & Te  & 38330  & 3373  & 3625  & 5949  & 305  & 181    & 12947  & 51458    \\ \hline 
$\geq$0.5 & Tr  & 170998 & 13620 & 14554 & 28823 & 1213 & 1385   & 56997  & 229380   \\
                     & Val & 42752  & 3405  & 3640  & 7205  & 303  & 346    & 14250  & 57348    \\
                     & Te  & 53439  & 4257  & 4548  & 9008  & 381  & 432    & 17813  & 71684    \\ \hline
$\geq$0.4 & Tr  & 228614 & 17651 & 18815 & 43710 & 1718 & 3354   & 80176  & 312144   \\
                     & Val & 57157  & 4412  & 4706  & 10926 & 430  & 838    & 20044  & 78039    \\
                     & Te  & 71444  & 5517  & 5879  & 13661 & 539  & 1047   & 25057  & 97548    \\ \hline
$\geq$0.3 & Tr  & 281952 & 23861 & 25277 & 57878 & 2624 & 5884   & 107016 & 394852   \\
                     & Val & 70492  & 5965  & 6322  & 14468 & 656  & 1470   & 26755  & 98717    \\
                     & Te  & 88113  & 7457  & 7899  & 18089 & 822  & 1837   & 33445  & 123395   \\ \hline
\end{tabular} 
}
\tablefoot{MG: Merger, GX: Galaxy, Tr: training, Te: test and Val: validation.}
\end{table}

\begin{table}[h!]
\centering
\caption{Results of the CNN and BCNN on several evaluation measures for each cumulative range. }
\label{res-ll-range}
\begin{tabular}{@{}clccccc@{}}
\hline \hline
\textbf{Agr.}   & \multicolumn{1}{c}{\textbf{Model}} & \textbf{EM} & \textbf{Accuracy} & \textbf{hR} & \textbf{hP} & \textbf{hF} \\ \hline
1   & CNN                                & 79.1        & 89.16             & 91.41       & 91.22       & 91.32       \\
                     & BCNN                               & 78.99       & 89.09             & 91.25       & 91.26       & 91.26       \\ \hline 
$\geq$0.9 & CNN                                & 83.73       & 91.55             & 93.06       & 93.03       & 93.04       \\
                     & BCNN                               & 83.36       & 91.31             & 92.71       & 92.99       & 92.85        \\ \hline 
$\geq$0.8 & CNN                                & 89.05       & 94.24             & 95.11       & 95.02       & 95.06       \\
                     & BCNN                               & 88.93       & 94.14             & 94.95       & 95.02       & 94.98        \\ \hline 
$\geq$0.7 & CNN                                & 91.62       & 95.5              & 96.09       & 96.01       & 96.05       \\
                     & BCNN                               & 91.39       & 95.31             & 95.89       & 95.89       & 95.89       \\ \hline
$\geq$0.6 & CNN                                & 92.14       & 95.66             & 96.14       & 96.17       & 96.16       \\
                     & BCNN                               & 92.18       & 95.65             & 96.14       & 96.14       & 96.14       \\ \hline 
$\geq$0.5 & CNN                                & 91.59       & 95.06             & 95.62       & 95.66       & 95.64       \\
                     & BCNN                               & 91.47       & 94.96             & 95.61       & 95.51       & 95.56        \\ \hline
$\geq$0.4 & CNN                                & 88.61       & 92.73             & 93.33       & 93.96       & 93.64       \\
                     & BCNN                               & 88.42       & 92.56             & 93.51       & 93.59       & 93.55       \\ \hline 
$\geq$0.3 & CNN                                & 84.39       & 89.62             & 90.46       & 91.46       & 90.96       \\
                     & BCNN                               & 84.04       & 89.32             & 90.54       & 91.03       & 90.78       \\ \hline
\end{tabular}
\tablefoot{The training is carried out only on the last layer of the CNN.}
\end{table}

\begin{table}[h!]
\centering
\caption{Results of the CNN and BCNN on several evaluation measures for each cumulative range. }
\label{res-al-range}
\begin{tabular}{@{}clccccc@{}}
\hline \hline
\textbf{Agr.}   & \multicolumn{1}{c}{\textbf{Model}} & \textbf{EM} & \textbf{Accuracy} & \textbf{hR} & \textbf{hP} & \textbf{hF} \\ \hline
1  & CNN                                & 99.01                           & 99.25                                 & 99.48                           & 99.31                           & 99.4                            \\
                           & BCNN                               & 98.96       & 99.17             & 99.46       & 99.21       & 99.34                           \\ \hline
$\geq$0.9 & CNN                                & 99.67                           & 99.76                                 & 99.85                           & 99.77                           & 99.81                           \\
                           & BCNN                               & 99.51       & 99.64             & 99.71       & 99.72       & 99.72                            \\ \hline
$\geq$0.8 & CNN                                & 99.55                           & 99.68                                 & 99.78                           & 99.7                            & 99.74                           \\
                           & BCNN                               & 99.53       & 99.65             & 99.72       & 99.72       & 99.72                           \\ \hline
$\geq$0.7 & CNN                                & 99.75                           & 99.81                                 & 99.83                           & 99.85                           & 99.84                           \\
                           & BCNN                               & 99.71       & 99.78             & 99.8        & 99.85       & 99.82                           \\ \hline
$\geq$0.6 & CNN                                & 99.69                           & 99.77                                 & 99.82                           & 99.79                           & 99.8                            \\
                           & BCNN                               & 99.67       & 99.76             & 99.8        & 99.79       & 99.79                           \\ \hline
$\geq$0.5 & CNN                                & 99.19                           & 99.41                                 & 99.5                            & 99.5                            & 99.5                            \\
                           & BCNN                               & 99.14       & 99.37             & 99.47       & 99.46       & 99.47                           \\ \hline
$\geq$0.4 & CNN                                & 96.98                           & 97.81                                 & 98.29                           & 97.99                           & 98.14                           \\
                           & BCNN                               & 96.7        & 97.61             & 98.19       & 97.78       & 97.98                           \\ \hline 
$\geq$0.3 & CNN                                & 93.14                           & 95.07                                 & 95.79                           & 95.79                           & 95.79                           \\
                           & BCNN                               & 92.84       & 94.84             & 95.76       & 95.48       & 95.62                           \\ \hline

\end{tabular}
\tablefoot{The training updates the weights of all the layers of the CNN.}
\end{table}

\FloatBarrier

\subsection{Number of images and other evaluation metrics:  exclusive ranges of agreement}\label{ranges}

\begin{small}
\begin{table}[h!]
\centering
\caption{Number of images associated to each exclusive range.}
\label{num-ranges}
\begin{tabular}{@{}ccccc@{}}
\hline
\textbf{Agr.} & \textbf{Training} & \textbf{Test} & \textbf{Validation} & \textbf{Total} \\ \hline
1                  & 6166              & 1928          & 1542                & 9636           \\
(1,0.9]          & 15244             & 4765          & 3812                & 23821          \\
(0.9,0.8]          & 36503             & 11408         & 9126                & 57037          \\
(0.8,0.7]          & 50118             & 15662         & 12530               & 78310          \\
(0.7,0.6]          & 57598             & 18000         & 14400               & 89998          \\
(0.6,0.5]          & 64964             & 20302         & 16241               & 101507         \\
(0.5,0.4]          & 83269             & 26022         & 20818               & 130109         \\
(0.4,0.3]          & 83614             & 26130         & 20904               & 130648         \\
(0.3,0.2]          & 28967             & 9053          & 7242                & 45262          \\ \hline
\end{tabular}
\end{table}

\begin{table}[h!]
\centering
\caption{Results of the CNN and BCNN on several evaluation measures for each exclusive range with transfer learning. }
\label{res-ranges}
\begin{tabular}{@{}clccccc@{}}
\hline \hline
\textbf{Agr.}         & \multicolumn{1}{c}{\textbf{Model}} & \textbf{EM} & \textbf{Accuracy} & \textbf{hR} & \textbf{hP} & \textbf{hF} \\ \hline
1  & CNN                                & 99.12       & 99.33             & 99.52       & 99.4        & 99.46       \\
                           & BCNN                               & 99.22       & 99.43             & 99.6        & 99.48       & 99.54       \\ \hline
(1,0.9] & CNN                                & 99.66       & 99.75             & 99.84       & 99.76       & 99.8        \\
                           & BCNN                               & 99.41       & 99.56             & 99.63       & 99.68       & 99.65        \\ \hline 
(0.9,0.8] & CNN                                & 99.7        & 99.77             & 99.84       & 99.78       & 99.81       \\
                           & BCNN                               & 99.66       & 99.75             & 99.76       & 99.82       & 99.79       \\ \hline
(0.8,0.7] & CNN                                & 99.39       & 99.55             & 99.54       & 99.67       & 99.6        \\
                           & BCNN                               & 99.32       & 99.5              & 99.49       & 99.65       & 99.57       \\ \hline
(0.7,0.6] & CNN                                & 99.43       & 99.59             & 99.65       & 99.66       & 99.65       \\
                           & BCNN                               & 99.34       & 99.53             & 99.59       & 99.61       & 99.6        \\ \hline 
(0.6,0.5] & CNN                                & 97.46       & 98.13             & 98.52       & 98.31       & 98.41       \\
                           & BCNN                               & 97.27       & 97.96             & 98.42       & 98.15       & 98.29       \\ \hline 
(0.5,0.4] & CNN                                & 90.3        & 92.95             & 94.4        & 93.74       & 94.07       \\
                           & BCNN                               & 89.85       & 92.63             & 94.28       & 93.38       & 93.83       \\ \hline
(0.4,0.3] & CNN                                & 78.06       & 84.24             & 87.57       & 86.19       & 86.88       \\
                           & BCNN                               & 78.02       & 84.14             & 87.26       & 86.36       & 86.81       \\ \hline
\end{tabular}
\end{table}
\end{small}

\subsection{Additional evaluation metrics for the evaluation on the full test set}\label{full}

\begin{small}
\begin{table}[h!]
\centering
\caption{Results of the CNN and BCNN trained on cumulative ranges on several evaluation measures assessed in the \textit{full} test set. }
\label{res-cum-all}
\begin{tabular}{@{}cllllll@{}}
\hline \hline
\textbf{Agr.}         & \multicolumn{1}{c}{\textbf{Model}} & \multicolumn{1}{c}{\textbf{EM}} & \multicolumn{1}{c}{\textbf{Accuracy}} & \multicolumn{1}{c}{\textbf{hR}} & \multicolumn{1}{c}{\textbf{hP}} & \multicolumn{1}{c}{\textbf{hF}} \\ \hline
1  & CNN                                & 60.12                           & 72.73                                 & 81.85                           & 73.38                           & 77.38                           \\
                           & BCNN                               & 59.33       & 72.11             & 81.45       & 72.85       & 76.91                           \\ \hline 
$\geq$0.9 & CNN                                & 64.36                           & 75.77                                 & 83.6                            & 76.24                           & 79.75                           \\
                           & BCNN                               & 61.75       & 74.02             & 82.37       & 74.99       & 78.51                           \\ \hline
$\geq$0.8 & CNN                                & 68.06                           & 78.28                                 & 85.17                           & 78.62                           & 81.77                           \\
                           & BCNN                               & 64.84       & 76.08             & 83.78       & 76.77       & 80.13                           \\ \hline
$\geq$0.7 & CNN                                & 71.71                           & 80.89                                 & 86.65                           & 81.25                           & 83.86                           \\
                           & BCNN                               & 68.89       & 78.9              & 85.51       & 79.38       & 82.33                           \\ \hline
$\geq$0.6 & CNN                                & 75.91                           & 83.92                                 & 88.12                           & 84.55                           & 86.3                            \\
                           & BCNN                               & 72.93       & 81.82             & 87.1        & 82.32       & 84.64                           \\ \hline 
$\geq$0.5 & CNN                                & 78.64                           & 85.81                                 & 89.05                           & 86.68                           & 87.85                           \\
                           & BCNN                               & 76.62       & 84.42             & 88.41       & 85.15       & 86.75                           \\ \hline
$\geq$0.4 & CNN                                & 81.16                           & 87.44                                 & 89.33                           & 88.99                           & 89.16                           \\
                           & BCNN                               & 79.99       & 86.64             & 89.21       & 87.93       & 88.56                           \\ \hline 
$\geq$0.3 & CNN                                & 81.95                           & 87.89                                 & 89.18                           & 89.83                           & 89.5                            \\
                           & BCNN                               & 81.44       & 87.48             & 89.18       & 89.31       & 89.24                           \\ \hline
\end{tabular}
\tablefoot{The training is carried out only on the last layer of the CNN.}
\end{table}

\begin{table}[h!]
\centering
\caption{Results of the CNN and BCNN trained on the cumulative ranges on several evaluation measures assessed in the \textit{full} test set. }
\label{res-al-all}
\begin{tabular}{@{}cllllll@{}}
\hline \hline
\textbf{Agr.}         & \multicolumn{1}{c}{\textbf{Model}} & \multicolumn{1}{c}{\textbf{EM}} & \multicolumn{1}{c}{\textbf{Accuracy}} & \multicolumn{1}{c}{\textbf{hR}} & \multicolumn{1}{c}{\textbf{hP}} & \multicolumn{1}{c}{\textbf{hF}} \\ \hline
1   & CNN                                & 68.9                            & 78.24                                 & 85.76                           & 78.24                           & 81.82                           \\
                           & BCNN                               & 61.28       & 72.88             & 82.53       & 73.3        & 77.64                            \\ \hline
$\geq$0.9 & CNN                                & 74.02                           & 81.57                                 & 88.16                           & 81.22                           & 84.55                           \\
                           & BCNN                               & 72.02       & 80.45             & 86.77       & 81.25       & 83.92                          \\ \hline
$\geq$0.8 & CNN                                & 79.9                            & 85.84                                 & 90.61                           & 85.66                           & 88.07                           \\
                           & BCNN                               & 78.42       & 84.77             & 89.71       & 85.04       & 87.32                           \\ \hline
$\geq$0.7 & CNN                                & 85.81                           & 90.04                                 & 92.58                           & 90.51                           & 91.53                           \\
                           & BCNN                               & 85.03       & 89.51             & 92.21       & 90.07       & 91.13                           \\ \hline 
$\geq$0.6 & CNN                                & 86.63                           & 90.57                                 & 93.25                           & 90.81                           & 92.01                           \\
                           & BCNN                               & 86.27       & 90.33             & 93.00          & 90.69       & 91.83                           \\ \hline 
$\geq$0.5 & CNN                                & 89.5                            & 92.53                                 & 94.12                           & 93.14                           & 93.63                           \\
                           & BCNN                               & 89.12       & 92.26             & 94.04       & 92.80        & 93.42                            \\ \hline 
$\geq$0.4 & CNN                                & 90.21                           & 93.03                                 & 94.69                           & 93.45                           & 94.07                           \\
                           & BCNN                               & 89.62       & 92.60              & 94.49       & 92.99       & 93.73                           \\ \hline
$\geq$0.3 & CNN                                & 90.91                           & 93.47                                 & 94.51                           & 94.38                           & 94.45                           \\
                           & BCNN                               & 90.51       & 93.17             & 94.43       & 94.00          & 94.22                               \\ \hline
\end{tabular}
\tablefoot{The training updates the weights of all the layers of the CNN.}
\end{table}

\begin{table}[h!]
\centering
\caption{Results of the CNN and BCNN trained by exclusive ranges on several evaluation measures assessed in the \textit{full} test set. }
\label{res-ranges-all}
\begin{tabular}{@{}clccccc@{}}
\hline \hline
\textbf{Agr.}         & \multicolumn{1}{c}{\textbf{Model}} & \textbf{EM} & \textbf{Accuracy} & \textbf{hR} & \textbf{hP} & \textbf{hF} \\ \hline
1  & CNN                                & 69.14       & 78.38             & 85.87       & 78.37       & 81.95       \\
                           & BCNN                               & 61.15       & 72.95             & 82.62       & 73.29       & 77.68        \\ \hline
(1,0.9] & CNN                                & 74.86       & 82.38             & 88.39       & 82.18       & 85.17       \\
                           & BCNN                               & 73.06       & 81.23             & 86.91       & 81.99       & 84.38       \\ \hline 
(0.9,0.8] & CNN                                & 82.55       & 87.74             & 91.65       & 87.7        & 89.63       \\
                           & BCNN                               &  80.81       & 86.43             & 90.59       & 86.84       & 88.68       \\ \hline 
(0.8,0.7] & CNN                                & 87.06       & 90.84             & 92.89       & 91.57       & 92.22       \\
                           & BCNN                               & 85.68       & 89.85             & 92.37       & 90.55       & 91.45       \\ \hline 
(0.7,0.6] & CNN                                & 88.42       & 91.82             & 93.67       & 92.43       & 93.05       \\
                           & BCNN                               & 87.94       & 91.50              & 93.5        & 92.08       & 92.78       \\ \hline 
(0.6,0.5] & CNN                                & 88.94       & 92.10              & 94.14       & 92.46       & 93.29       \\
                           & BCNN                               & 88.47       & 91.76             & 93.96       & 92.12       & 93.03       \\ \hline 
(0.5,0.4] & CNN                                & 90.72       & 93.32             & 94.56       & 94.07       & 94.31       \\
                           & BCNN                               & 90.38       & 93.08             & 94.46       & 93.79       & 94.12       \\ \hline 
(0.4,0.3] & CNN                                & 90.51       & 93.17             & 94.50        & 93.9        & 94.20        \\
                           & BCNN                               & 90.39       & 93.03             & 94.29       & 93.94       & 94.12       \\ \hline
\end{tabular}
\tablefoot{The training updates the weights of all the layers of the CNN, and transfer learning is carried out.}
\end{table}
\end{small}

\FloatBarrier

\onecolumn

\section{Sample images}\label{sec:si}

\begin{figure}[tbh]
\centering
\begin{tabular}{cccc}
GALAXY-SPIRAL-ACW & GALAXY-SPIRAL-EDGE & GALAXY-SPIRAL-ACW  & GALAXY-SPIRAL-ACW\\ 
\includegraphics[width=4cm]{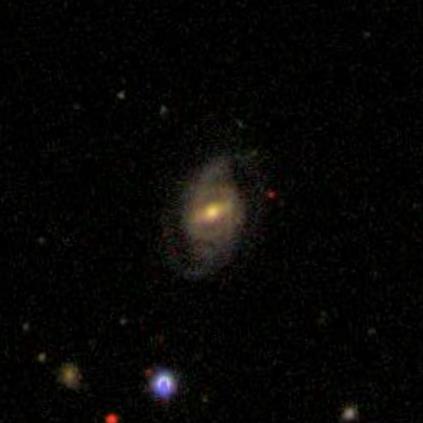} & \includegraphics[width=4cm]{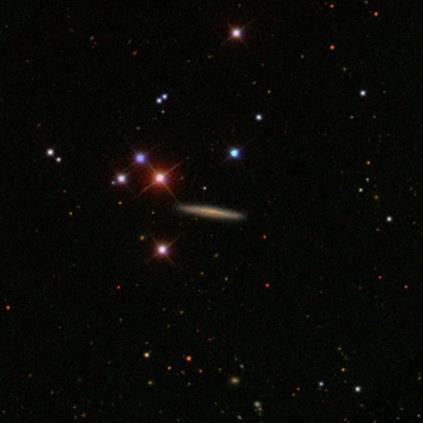} & \includegraphics[width=4cm]{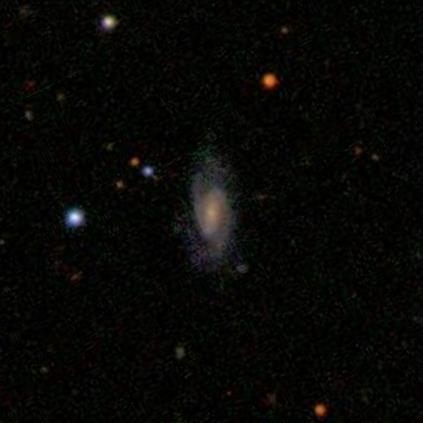} &  \includegraphics[width=4cm]{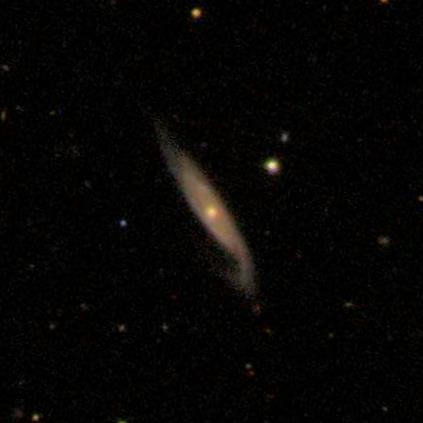}\\
GALAXY-SPIRAL-CW & DK & GALAXY-SPIRAL-CW & GALAXY-SPIRAL-EDGE \\ 
\end{tabular}
\caption{Example images in which the ML differs from the crowd for $agr = 1$. On top the annotations by the crowd and below the ML classification.}
\label{img1}
\end{figure}

\begin{figure}[tbh]
\centering
\begin{tabular}{cccc}
GALAXY-SPIRAL-EDGE & GALAXY-ELLIPTICAL & GALAXY-ELLIPTICAL & GALAXY-ELLIPTICAL\\ 
\includegraphics[width=4cm]{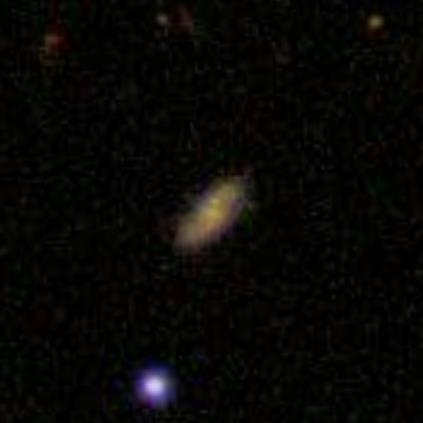} & \includegraphics[width=4cm]{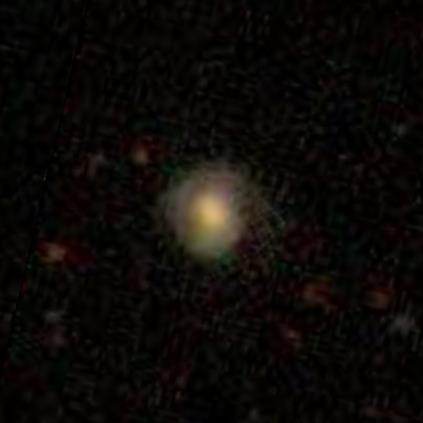} & \includegraphics[width=4cm]{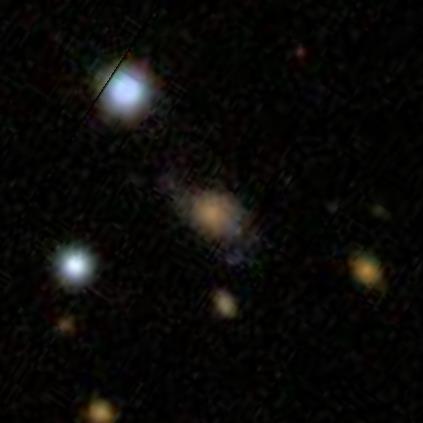} & \includegraphics[width=4cm]{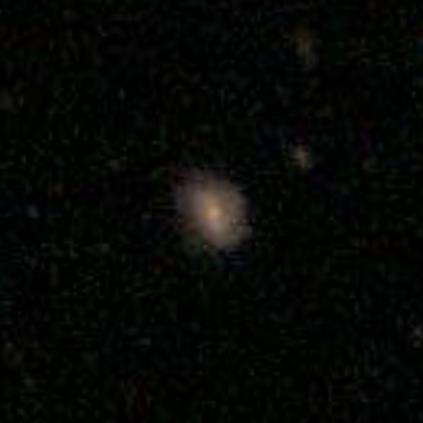}\\
GALAXY-SPIRAL-ACW & GALAXY-SPIRAL-ACW & GALAXY-SPIRAL-ACW & GALAXY-SPIRAL-CW \\ 
\\
GALAXY-ELLIPTICAL & GALAXY-ELLIPTICAL &  GALAXY-ELLIPTICAL & DK\\ 
\includegraphics[width=4cm]{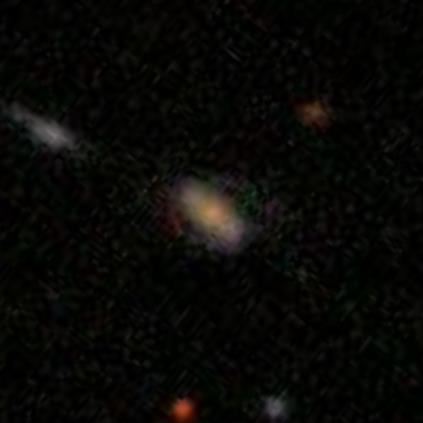} & \includegraphics[width=4cm]{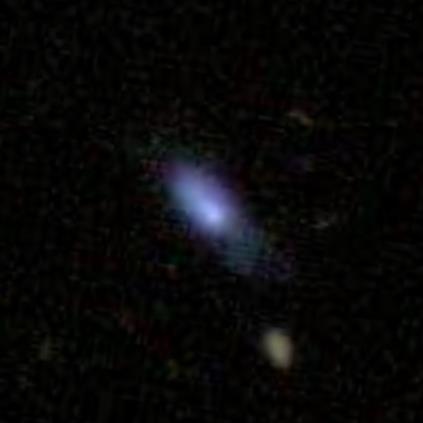} &  
\includegraphics[width=4cm]{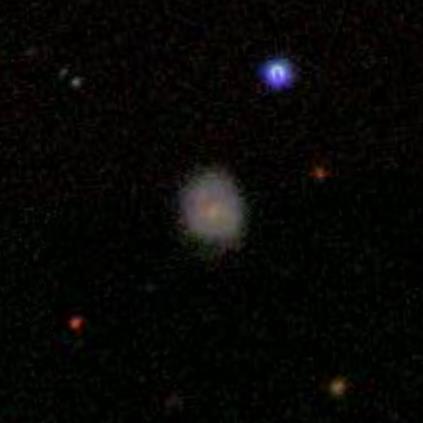} & \includegraphics[width=4cm]{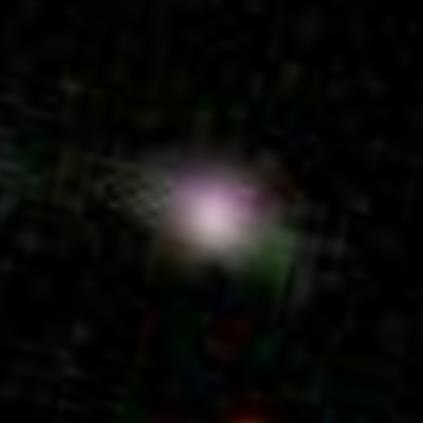}\\
GALAXY-SPIRAL-EDGE & GALAXY-SPIRAL-EDGE &  GALAXY-SPIRAL-ACW & GALAXY-ELLIPTICAL \\
\end{tabular}
\caption{Example images in which the ML differs from the crowd for difficult cases, $agr \leq 0.4$. On top the annotations by the crowd and below the ML classification.}
\label{img2}
\end{figure}

In this appendix, we show some sample images in which the CNN classification differs from the crowd. We include 4 examples of perfect agreement, Figure \ref{img1} (there are only 12 cases in which they differ out of 1837 test images); and 8 examples of $agr \leq 0.4$, Figure \ref{img2}. 

\clearpage

\end{appendix}
\end{document}